\documentclass[nofootinbib, aps,prb,twocolumn,superscriptaddress]{revtex4-2}

\usepackage[hidelinks]{hyperref}

\usepackage{physics}
\usepackage{balance}
\usepackage{amssymb}
\usepackage{suffix}
\usepackage{mathtools}
\usepackage[utf8]{inputenc}
\usepackage{booktabs}
\usepackage{cases}
\usepackage[multiple]{footmisc}
\usepackage{dcolumn}
\usepackage{color,soul}
\usepackage{rotating}
\usepackage{perpage}
\usepackage{siunitx}
\usepackage{xcolor}
\usepackage{soul}
\usepackage{amsmath}
\usepackage{tikz}
\usepackage[T1]{fontenc}
\usepackage{etoolbox}
\usepackage{graphics}
\usepackage{siunitx}
\usepackage{float}	
\usepackage{collref}
\usepackage{multirow}
\usepackage{mathtools}
\usepackage{bm}
\usepackage{url}

\usepackage{tikz}
\usepackage{tikz-3dplot}
\usepackage{accents}

\makeatletter
\newcommand{\doublewidetilde}[1]{{%
  \mathpalette\double@widetilde{#1}%
}}
\newcommand{\double@widetilde}[2]{%
  \sbox\z@{$\m@th#1\widetilde{#2}$}%
  \ht\z@=.9\ht\z@
  \widetilde{\box\z@}%
}
\makeatother

\usepackage{etoolbox,lipsum}

\begin{document}

\title{Controlling transport with structured long-range hopping}

 \author{Boris Gurevich}
 \email{Contact author: boris.gurevich@utdallas.edu}
 \address{Department of Physics, The University of Texas at Dallas, Richardson, Texas 75080, USA}
 \author{Michael H. Kolodrubetz}
 \address{Department of Physics, The University of Texas at Dallas, Richardson, Texas 75080, USA}

\date{\today}

\begin{abstract}
The combined effects of disorder and long-range hopping play an important role in determining localization and transport in low-dimensional quantum systems.
We study transport in a one-dimensional disordered chain with nearest-neighbor hopping supplemented by a long-range hopping term. The long-range coupling is characterized by site-dependent amplitudes that may be either uniform or random. We investigate the interplay between diagonal disorder and long-range hopping, characterizing the resulting transport efficiency. We find that nearest-neighbor hopping in the presence of uniform long-range hopping amplitudes strongly suppresses delocalization over a range of weak disorder. The disorder-enhanced transport regime, established in previous studies, can be understood here as the recovery of delocalization from this strongly suppressed regime. An analytical solution in the Anderson eigenstate basis reveals the microscopic origin of this behavior. In particular, we show that the low-disorder regime of strongly suppressed delocalization can be eliminated by introducing randomness into the long-range hopping amplitudes, providing an additional mechanism for enhancing transport and qualitatively changing the response of the system to disorder.
\end{abstract}

\maketitle

\section{Introduction}

Enhancing transport efficiency in molecular and nanoscale materials is important for efficient transfer of electronic and optical excitations. 
In particular, efficient exciton transport can enable energy transfer over extended distances while reducing the impact of localization and energy losses~\cite{Brixner2017,Mikhnenko2015}.

Exciton transport is also generally accompanied by the spreading of quantum information encoded in the excitonic state, making efficient and controllable transport relevant to platforms for quantum information processing~\cite{10.1021/acs.jpclett.4c00180, wasielewski2020exploiting, sarovar2010quantum}. Identifying mechanisms that enhance transport in disordered materials may therefore provide new ways to control both energy and quantum-information transfer.

Disorder-induced localization is a central phenomenon in quantum transport~\cite{Anderson1958, Abrahams1979}.
For a one-dimensional Anderson chain, increasing on-site disorder suppresses
the spatial extent of the eigenstates.
Long-range hopping can qualitatively
modify this behavior. Previous studies~\cite{Chavez2021, mqw1-jfvb, wu2024dit} have demonstrated the emergence of a
disorder-enhanced transport (DET) regime in one-dimensional disordered
systems supplemented by long-range hopping induced by cavity quantum
electrodynamics (cavity QED)~\cite{Allard2022}. In these systems, increasing disorder
can enhance transport over an intermediate range of disorder strengths,
contrary to the conventional picture of disorder-induced localization.

Our work focuses on the related phenomenon of a strong suppression of transport induced by nearest-neighbor hopping over a range of weak disorder. We demonstrate that nearest-neighbor hopping can give rise to a suppressed spatial variance (SSV) regime. The disorder-enhanced transport regime can then be understood as the recovery of the spatial variance from this SSV regime. We further show that the SSV effect can be reduced or even nearly eliminated by controlling the parameters of the long-range interaction.

The remainder of this paper is organized as follows. Section~\ref{sec_model} reviews the model and introduces the key quantities used throughout this study, including the inverse participation ratio and spatial variance. Section~\ref{sec_numeric} presents the numerical results and discusses the observed behavior. In Sections~\ref{sec_richardson} and~\ref{sec_anderson}, we formulate the solution of the model in the Anderson eigenstate basis and derive analytical relations connecting the statistical properties of the system. Sections~\ref{sec_test_num} and~\ref{sec_ssv} analyze the numerical and analytical results and identify the mechanisms responsible for the formation of the SSV regime. Section~\ref{sec_period} investigates the localization mechanism in the presence of periodically structured long-range hopping amplitudes. Section~\ref{sec_current} examines transport efficiency in terms of the effective current. Finally, Section~\ref{sec_outlook} summarizes our results, discusses their physical implications, and outlines directions for future research.

\section{Model}
\label{sec_model}

\begin{figure}[t]
\centering
\includegraphics[width=\columnwidth]{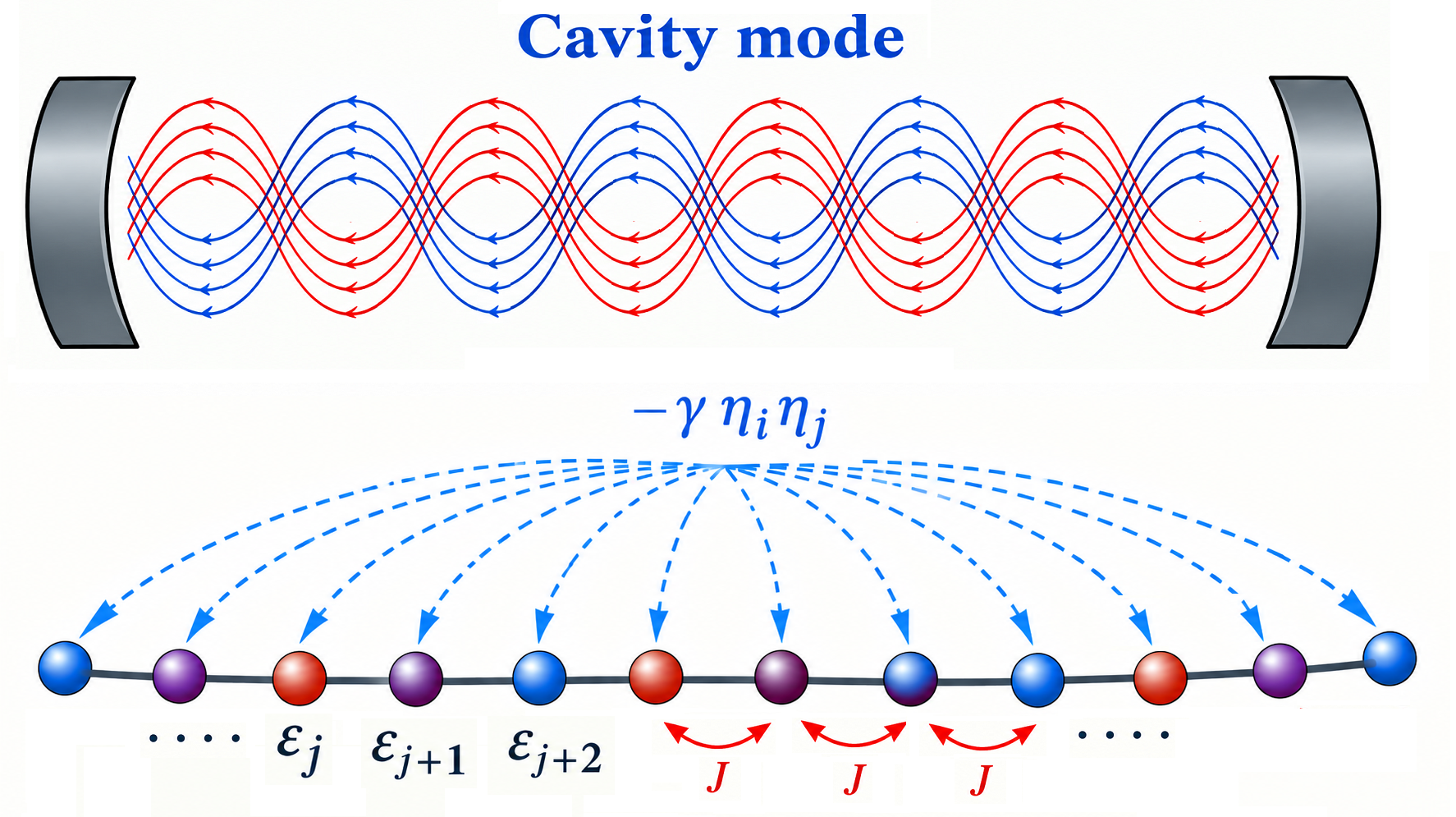}
\caption{Schematic illustration of the one-dimensional chain model described by $H = H_A + H_L$ in Eqs.~(\ref{eq:HT},\ref{eq:HA},\ref{eq:HL}). The on-site potentials are uniformly distributed according to $\epsilon_j\in[-W/2,W/2]$, where $W$ controls the disorder strength. In addition to nearest-neighbor hopping with amplitude $J$, the system exhibits long-range hopping induced by the cavity mode, with matrix elements $-\gamma\eta_i\eta_j$ coupling all pairs of sites. For a standing-wave cavity, the coupling of an emitter at site $j$ depends on its position within the cavity-mode profile and can be written as $t_j=t_0\eta_j$, where $t_0$ is the peak coupling strength and $\eta_j=\cos(kx_j)$ is the site-dependent mode amplitude.
}
\label{fig:model}
\end{figure}

We consider a tight-binding model with a one-dimensional closed chain of $N$ sites,
described by the Hamiltonian
\begin{equation}
H=H_A+H_L,
\label{eq:HT}
\end{equation}
with periodic boundary conditions, where
\begin{equation}
H_A=
-J\sum_{j}
\left(
|j\rangle\langle j+1|
+ h.c. \right)
+\sum_{j}\epsilon_j|j\rangle\langle j|
\label{eq:HA}
\end{equation}
is the one-dimensional Anderson model~\cite{Anderson1958}, and
\begin{equation}
H_L=-\gamma\sum_{i,j}\eta_i\eta_j|i\rangle\langle j|
\label{eq:HL}
\end{equation}
is a long-range hopping term obtained by integrating out an off-resonant cavity mode, as depicted in Fig.~\ref{fig:model}. 
Unless stated otherwise, we set the nearest-neighbor hopping amplitude to $J = 1$. Note that, for $J=0$, the model has factorizable all-to-all hopping and thus is a generalized Richardson model~\cite{Richardson1963, RichardsonSherman1964}.

The on-site energies are independent random variables uniformly distributed
according to
\begin{equation}
\epsilon_j\in[-W/2,W/2],
\label{eq:disorder}
\end{equation}
where $W$ controls the disorder strength. The overall strength of the
long-range coupling is set by $\gamma$, while $\eta_j$ determines its
spatial profile. We consider three cases. The first corresponds to uniform all-to-all long-range hopping with amplitude $\gamma$:
\begin{equation}
\eta_j=1.
\label{eq:eta_case1}
\end{equation}
The other two cases describe disorder in the long-range hopping amplitudes, with $\eta_j$ assigned uniformly at random within the respective ranges:
\begin{equation}
\eta_j\in[0,1],\qquad
\eta_j\in[-1,1].
\label{eq:eta_cases23}
\end{equation}

We now define the key quantities used throughout this work to characterize the localization and transport properties of the system.

First, we use the inverse participation ratio,
\begin{equation}
{\rm IPR}=\sum_j|\psi(j)|^4\,,
\label{eq:ipr}
\end{equation}
which is a proxy for the inverse of the number of occupied sites.

Second, we use the spatial variance. Given our choice of periodic boundary conditions, some care must be taken to define this variance.
For a normalized state with site probabilities $p(j) = |\psi(j)|^2$, we choose the definition
\begin{equation}
\sigma^2=
\frac{1}{2}
\sum_{i,j}
p(i)p(j)\,
{\rm dist}^2(i,j),
\label{eq:sigma}
\end{equation}
where ${\rm dist}(i,j)$ is the shortest-path distance on the ring.
\begin{equation}
\operatorname{dist}(i,j)=
\begin{cases}
|j-i|,& |j-i|\le N/2,\\
N-|j-i|,& |j-i|>N/2.
\end{cases}
\label{eq:dist}
\end{equation}

 This definition does not require a choice of a central position and therefore remains unambiguous on a closed topology. For an ordinary open geometry with Euclidean distance \( \operatorname{dist} (x, y) = | x - y | \) it reduces to the conventional variance:
\begin{equation}
\sigma^2 \ = \langle \psi | (x - x_{\rm c} )^2 | \psi \rangle  \label{eq:sigm_class}
\end{equation}
with the central point \( 
x_{\rm c} = \langle \psi | x | \psi \rangle
\).

For $N\gg1$, the average squared shortest-path distance over pairs of sites
is
\begin{equation}
\operatorname{dist}_{\rm av}^2
\simeq
\left(\frac{N}{2}\right)^2\int_0^1x^2\,dx
=\frac{N^2}{12}.
\label{eq:distav}
\end{equation}
A completely extended state has $p(j)=1/N$, for which the spatial variance is
\begin{equation}
\sigma_{\rm ex}^2
=\frac{1}{2}\operatorname{dist}_{\rm av}^2
=\frac{N^2}{24}.
\label{eq:sigmaex}
\end{equation}

When necessary, we distinguish averages over disorder realizations,
denoted by $\langle\cdots\rangle_{\rm dis}$, from averages over eigenstates.
For non-uniform $\eta_j$, the disorder average includes the possible
realizations of the random $\eta_j$ values in addition to the realizations
of the on-site disorder. Unless otherwise specified, $\langle\sigma^2\rangle$
and $\langle {\rm IPR}\rangle$ denote averages over all eigenstates and
disorder realizations.

\section{Numerical results}
\label{sec_numeric}

\begin{figure*}[t]
\centering
\includegraphics[width=\textwidth]{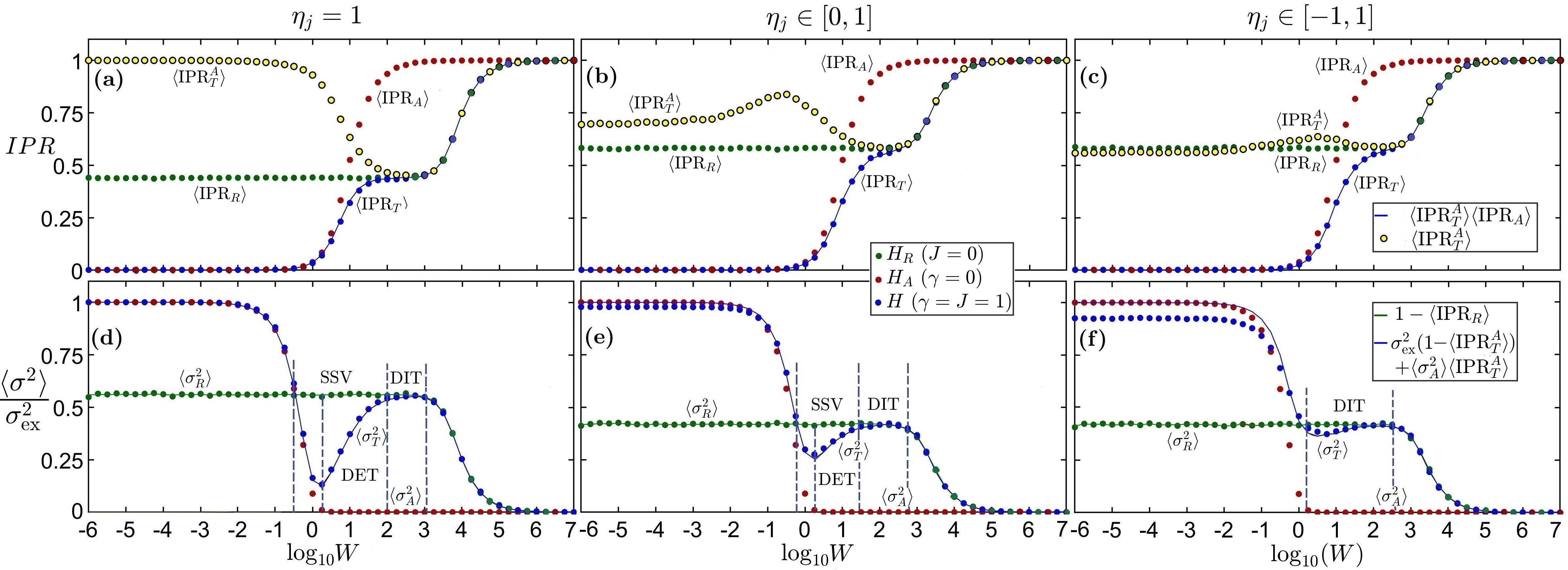}
\caption{System response to disorder over the wide range of disorder strengths $W \in [10^{-6}, 10^{7}]$, characterized by the ${\rm IPR}$ 
and spatial variance
$\langle \sigma^2 \rangle$. 
For each observable, we show the results applicable to the complete Hamiltonian $H$ (blue dots) and its Richardson $J=0$ (green dots) and Anderson $\gamma=0$ (red dots) limits. 
Top panels also show also show ${\rm IPR}^A_T$ (yellow rounded dots), which characterizes the localization of the complete Hamiltonian $H$ in the Anderson eigenbasis, together with the analytical results from Eq.~(\ref{eq:ipr_T_stat}), corresponding to ${\rm IPR}_T$ (blue line). 
Bottom panels also include
the analytical results from Eq.~(\ref{eq:richipr}), corresponding to the $\langle \sigma^2_R \rangle$ quantity (green line), and Eq.~(\ref{eq:sigma_T_stat}), corresponding to the $\langle \sigma^2_T \rangle$ quantity (blue line). All numerical calculations are performed for a system size of $N=1000$ and long-range coupling strength $\gamma=1$. Averages are taken over all eigenstates and nine disorder realizations. Error bars are smaller than the marker size.
}
\label{fig:comb_ipr_sgm}
\end{figure*}

Fig.~\ref{fig:comb_ipr_sgm} shows the average ${\rm IPR}$ and spatial variance over a wide range of disorder strengths $W$ for the full Hamiltonian $H$ and its limiting cases, $H_R$ ($J=0$) and $H_A$ ($\gamma=0$). The quantities associated with the full
Hamiltonian are labeled ${\rm IPR}_T$ and $\sigma_T^2$ (T for Total), those
for $J=0$ are labeled ${\rm IPR}_R$ and $\sigma_R^2$ (R for Richardson), and
those for $\gamma=0$ are labeled ${\rm IPR}_A$ and $\sigma_A^2$ (A for
Anderson).

For all three $\eta_j$ configurations, the inverse participation ratio of the complete system, ${\rm IPR}_T$, shown in Fig.~\ref{fig:comb_ipr_sgm}(a)--(c), exhibits a similar dependence on $W$. At weak disorder, $W\to0$, the nearest-neighbor hopping dominates the spatial structure of the system, leading to extended, Bloch-like eigenstates. Consequently, ${\rm IPR}_T$ follows the inverse participation ratio of the Anderson model, ${\rm IPR}_A$, corresponding to the $\gamma=0$ limit. In the opposite limit, $W\to\infty$, the Anderson eigenstates become localized around individual sites, making the nearest-neighbor hopping  irrelevant. The long-range hopping then dominates the system dynamics, and ${\rm IPR}_T$ approaches the inverse participation ratio of the Richardson model, ${\rm IPR}_R$, corresponding to the $J=0$ limit. At high disorder strength, ${\rm IPR}_T$ approaches unity, indicating strong localization.
These observations suggest that, at a certain disorder strength, the long-range hopping becomes dominant, and the eigenstates of $H_T$ increasingly resemble those of $H_R$.

On the other hand, the nonmonotonic behavior of the spatial variance
[Fig.~\ref{fig:comb_ipr_sgm}(d, e, f)] provides more information about
the intermediate transport regimes than the ${\rm IPR}$ alone, as the ${\rm IPR}$ only tells us about how many sites are occupied, not whether these states are in a localized region. For the uniform case $\eta_j=1$ [Fig.~\ref{fig:comb_ipr_sgm}(d)], the spatial variance
is strongly suppressed over a range of disorder strengths, where
$\sigma_T^2$ falls below the $\sigma_R^2$ level. The subsequent recovery
of $\sigma_T^2$ from this suppressed spatial variance (SSV) regime gives rise to the
disorder-enhanced transport (DET) regime. At disorder strengths above
the SSV regime,
$\sigma_T^2$ remains close to $\sigma_R^2$, producing the
disorder-independent transport (DIT) plateau until both $\sigma_T^2$ and $\sigma_R^2$ collapse to zero showing strong localization above certain level of disorder strength.

The increase in the average $\sigma^2$ during the recovery from the SSV regime, accompanied by an increase in the ${\rm IPR}$, may seem counterintuitive at first but has a simple explanation. When nearest-neighbor hopping dominates, a particle or excitation is more likely to hop to a nearby site at distance $1$ than to reach sites at distances of order $N/2$ through the long-range hopping channel, resulting in a strong suppression of the spatial variance compared to the long-range (Richardson) case. As the long-range hopping becomes dominant at higher disorder strengths, the wave packet can become concentrated on fewer sites while extending over larger distances, thereby increasing $\sigma^2$.

At the same time, introducing randomness into the long-range hopping amplitudes while retaining a positive bias, $\eta_j\in[0,1]$ [Fig.~\ref{fig:comb_ipr_sgm}(e)], significantly weakens the SSV effect. For the random profile $\eta_j\in[-1,1]$ [Fig.~\ref{fig:comb_ipr_sgm}(f)], both the SSV and DET regimes are nearly eliminated. Consequently, the system remains effectively in the DIT regime over the range of disorder strengths for which $\langle \sigma^2_A \rangle < \langle \sigma^2_R \rangle$ and the disorder remains below the critical strength established for the $J=0$ (Richardson) limit in Refs.~\cite{Modak2016,Yuzbashyan2005finitesize,celardo2016shielding}.

These observations show that the behavior cannot be understood solely from statistical properties such as the disorder strength or average long-range hopping amplitude. A microscopic solution is therefore required to identify the origin of the SSV regime and its sensitivity to the distribution of the long-range hopping amplitudes.

\section{The $J=0$ limit: Richardson model}
\label{sec_richardson}

Before considering the complete problem, it is useful to examine the
limit $J=0$. In this limit, the Hamiltonian consists of diagonal disorder
and the factorizable  long-range hopping term which corresponds to Richardson model~\cite{Richardson1963,RichardsonSherman1964}. 
The $J=0$ limit of $H$ is exactly solvable~\cite{Modak2016}.
The corresponding
eigenstates have the unnormalized site amplitudes
\begin{equation}
\psi^R_{\lambda}(j)=\frac{\eta_j}{\lambda-\epsilon_j},
\label{eq:richsol}
\end{equation}
where eigenvalue $\lambda$ is determined by the secular equation
\begin{equation}
\sum_j\frac{\eta_j^2}{\lambda-\epsilon_j}
=-\frac{1}{\gamma}.
\label{eq:richsec}
\end{equation}

The site probability $p^R_{\lambda}(j) \equiv |\psi^R(j)|^2$ depends on the local disorder potential
and on $\eta_j$, but has no explicit dependence on the spatial position
of the site. Consequently, after averaging over statistically
independent disorder realizations, the squared distance in
Eq.~(\ref{eq:sigma}) can be replaced by its average value,
$\operatorname{dist}_{\rm av}^2$, for any pair of distinct sites
$i\ne j$. For $i=j$, the distance is zero. Using the normalization
condition $\sum_j p^R_{\lambda}(j) = 1$, the spatial variance
can therefore be written as
\begin{align}
\langle \sigma_R^2 \rangle
&=
\frac{\operatorname{dist}_{\rm av}^2}{2}
\langle
\sum_{i\ne j} p^R_{\lambda}(i) \, p^R_{\lambda}(j)
\rangle
\nonumber\\
&=
\frac{\operatorname{dist}_{\rm av}^2}{2}
\langle
1-\sum_jp^R_{\lambda}(j)^2
\rangle
\nonumber\\
&=
\sigma_{\rm ex}^2
\left(1-\langle {\rm IPR}_R\rangle\right).
\label{eq:richipr}
\end{align}
Equation~(\ref{eq:richipr}) establishes a direct relation between the
spatial variance and the inverse participation ratio in the $J=0$
limit. The green curves in Fig.~\ref{fig:comb_ipr_sgm}(d,e,f), labeled
as $1-\langle {\rm IPR}_R\rangle$, provide a numerical test of this relation relation. Their dependence on $W$ closely follows that of
$\langle\sigma_R^2\rangle$, as expected from Eq.~(\ref{eq:richipr}).

Previous analyses of this model~\cite{Modak2016,Yuzbashyan2005finitesize,celardo2016shielding} show that the average participation ratio of the excited states remains approximately constant for disorder strengths below the characteristic scale $W_{\rm cr,R} \sim \gamma N\ln N$, assuming that $\eta_j^2$ values are of order unity. Within the approximation of equally spaced site energies, the low-disorder value is $\langle PR\rangle\approx 3$.

In our numerical calculations,
the average ${\rm IPR}$ in the low-disorder regime is approximately
$\langle {\rm IPR}_R\rangle\approx0.44$ for the uniform case
$\eta_j=1$, and $\langle {\rm IPR}_R\rangle\approx0.58$ for both random-amplitude
configurations, $\eta_j\in[0,1]$ and $\eta_j\in[-1,1]$.
According to Eq.~(\ref{eq:richipr}), these values correspond to
spatial variance plateaus of approximately
\begin{equation}
\langle\sigma_R^2\rangle
\approx 0.56\,\sigma_{\rm ex}^2
\qquad
(\eta_j=1),
\end{equation}
and
\begin{equation}
\langle\sigma_R^2\rangle
\approx 0.42\,\sigma_{\rm ex}^2
\qquad
(\eta_j\in[0,1]\,\, {\rm or} \,\,[-1,1]).
\end{equation}

Thus, if our earlier assumption that, above a certain disorder strength, the eigenstates of $H_T$ closely resemble those of $H_R$ holds, the low-disorder value of $\langle {\rm IPR}_R\rangle$ in the $J=0$ limit is expected to determine the corresponding spatial variance of the complete system $H_T$ in the disorder-independent transport (DIT) regime through Eq.~(\ref{eq:richipr}).

\section{Analytical solution in the Anderson eigenbasis}
\label{sec_anderson}

We now express the complete Hamiltonian $H=H_A+H_L$ in the
Anderson eigenbasis, defined by
\begin{equation}
H_A|\psi_n^A\rangle=\mu_n|\psi_n^A\rangle\,.
\label{eq:Anderson}
\end{equation}

The long-range matrix element in this basis is
\begin{align}
\langle\psi_n^A|H_L|\psi_m^A\rangle
&=
-\gamma
\sum_{i,j}\eta_i\eta_j
\langle\psi_n^A|i\rangle
\langle j|\psi_m^A\rangle
\nonumber\\
&=
-\gamma f_n^*f_m,
\label{eq:HLAnderson}
\end{align}
where \footnote[1]{Note that, as the Hamiltonian is real, all of the $\psi_n^A(j)$ values, and thus $f_n$, can be chosen to be real. However, we leave our solution in the more general form to allow simpler plane wave approximations.}
\begin{equation}
f_n\equiv\sum_j\eta_j\psi_n^A(j).
\label{eq:fn}
\end{equation}
Therefore, the full Hamiltonian takes the form
\begin{equation}
H=
\sum_n\mu_n|\psi_n^A\rangle\langle\psi_n^A|
-\gamma
\sum_{n,m}f_n^*f_m
|\psi_n^A\rangle\langle\psi_m^A|.
\label{eq:HAnderson}
\end{equation}

In the Anderson basis, the total Hamiltonian takes precisely the form of the Richardson model, with the caveat that $\mu_n$ and $f_n$ may be correlated, as discussed in Sections~\ref{sec_ssv} and~\ref{sec_period} below. Thus, the exact solution of the Richardson model can be applied to the present Hamiltonian. We expand an eigenstate of the Hamiltonian $H$ as

\begin{equation}
|\psi_k^T\rangle
=
\sum_n A_{nk}|\psi_n^A\rangle,
\label{eq:expansion}
\end{equation}
with the matrix elements $A_{nk}$ given by
\begin{equation}
A_{nk}
=
\langle\psi_n^A|\psi_k^T\rangle
=
\frac{\gamma f_n^*S_k}{\lambda_k-\mu_n},
\label{eq:Ank}
\end{equation}
where we have explicitly written the normalization factor $S_k$. Here, $\lambda_k$ denotes the $k$th
eigenvalue obtained from the secular equation
\begin{equation}
\sum_n\frac{|f_n|^2}{\lambda-\mu_n}
=-\frac{1}{\gamma}.
\label{eq:sec_anderson_bas}
\end{equation}

The $j$th site probability of the $k$th eigenstate of the full
Hamiltonian is
\begin{align}
p^T_k(j)
&=
\left|
\sum_n A_{nk}\psi_n^A(j)
\right|^2
\nonumber\\
&=
\sum_n |A_{nk}|^2|\psi_n^A(j)|^2+Q_k(j),
\label{eq:pk}
\end{align}
where
\begin{equation}
Q_k(j)=
\sum_{n\ne m}
A_{nk}^*A_{mk}
\psi_n^{A*}(j)\psi_m^A(j)
\label{eq:Q}
\end{equation}
is the interference term between eigenstates. 
As $N \to \infty$, arbitrarily small disorder causes Anderson localization, meaning that the tails of two randomly chosen states will not overlap. For finite $N$, as the Anderson wave functions have oscillatory tails, we assume that the contributions to $Q_k(j)$ have approximately equal
likelihood of being positive or negative. Based on this reasoning, the
terms contributing to the spatial variance and ${\rm IPR}$ that are linear in
$Q_k(j)$ are expected to largely cancel upon averaging over disorder
realizations, while contributions of order $|Q_k(j)|^2$ are expected to
remain small in the disorder range associated with the SSV regime and
above, where the Anderson eigenstates are already substantially
localized. Therefore, we introduce the following approximation for
$p_k^T(j)$:
\begin{equation}
p_k^T(j) \approx
\sum_n |A_{nk}|^2 \, p_n^A(j), \qquad
p_n^A(j) \equiv |\psi_n^A(j)|^2,
\label{eq:pk_approx}
\end{equation}
effectively neglecting the interference term.

In terms of this solution, both ${\rm IPR}_T$ and $\sigma^2_T$ can be expressed in the general form
\begin{equation}
V_{\rm gen} =
\frac{1}{N}\left\langle \sum_{n,k,m}
|A_{nk}|^2|A_{nm}|^2G(k,m) \right\rangle_{\rm dis},
\label{eq:sum_a_nm_form}
\end{equation}
where the average is taken over all eigenstates and an ensemble of disorder realizations. Here, $G(k,m)$ is a generalized distance between two Anderson eigenstates, defined as
\begin{equation}
G(k,m) \equiv
\sum_{j,i} p_k^A(j)p_m^A(i)g(j,i),
\label{eq:gen_dist_km}
\end{equation}
where $g(j,i)$ is a function of the site indices chosen according to the quantity of interest. When $g(i,j)$ is chosen as the Kronecker delta,
\begin{equation}
g(i,j)=\delta_{ij},
\end{equation}
Eq.~(\ref{eq:sum_a_nm_form}) yields the average ${\rm IPR}$ of the complete system. In contrast, when $g(i,j)$ is chosen as one half of the squared shortest-path distance between sites,
\begin{equation}
g(i,j)=\frac{1}{2}\operatorname{dist}^2(i,j),
\end{equation}
Eq.~(\ref{eq:sum_a_nm_form}) gives the average spatial variance.

We also note an important similarity between the statistical properties of the solution for $H$ in the Anderson eigenbasis and those of the $J=0$ limit in the coordinate basis discussed in Section~\ref{sec_richardson}. Just as the Richardson wave functions $\psi^R_{\lambda}(j)$ depend only on the local disorder potential $\epsilon_j$ and on $\eta_j$, without explicit dependence on the spatial position of the site, the coefficients $A_{nk}$ depend only on the Anderson eigenvalues $\mu_n$ and the coefficients $f_n$, and not directly on the spatial shapes of the Anderson eigenstates $|\psi_n^A\rangle$. The coefficients $f_n$ are determined by the the sum of the $\eta_j\psi_n^A(j)$ terms as defined in Eq.~(\ref{eq:fn}), which contains information about the Anderson eigenstate. However, this summation is not expected to induce a strong correlation between $f_n$ and the spatial structure of $\psi_n^A(j)$. In general, a large number of Anderson eigenstates with very different spatial profiles can produce similar  values of $f_n$. Thus, while $f_n$ is determined by the eigenstate amplitudes, it does not uniquely characterize their spatial structure.
The only exception is the fully symmetric ground state, $\psi(j)=1/\sqrt{N}$, in the low-disorder limit $W\to0$. However, because of its symmetry, this state does not encode a preferred spatial position and remains invariant under any permutation of the site indices $j\leftrightarrow i$.

Based on this reasoning, we expect the terms involving the coefficients $A_{nk}$ in Eq.~(\ref{eq:sum_a_nm_form}) to be largely uncorrelated with $G(k,m)$, 
which is determined by the spatial profiles of the Anderson eigenstates. This allows us to statistically separate these factors when performing the disorder average.
At the same time, the generalized distance $G(k,m)$ is expected to exhibit different statistical properties depending on whether $k=m$ or $k\ne m$. 
In particular, $G(k,k)$ characterizes the spatial structure of a single Anderson eigenstate, whereas $G(k,m)$ for $k\ne m$ characterizes the distance between two different Anderson eigenstates.
We therefore treat the corresponding contributions, $|A_{nk}|^4G(k,k)$ and $|A_{nk}|^2|A_{nm}|^2G(k,m)$, separately when performing the disorder average. 
This separation is analogous to the treatment of the $J=0$ Richardson case discussed in Section~\ref{sec_richardson}, where the ensemble-averaged distance 
between two sites had to be evaluated separately for the cases $i=j$ and $i\ne j$.

By statistically separating the $A_{nk}$-dependent factors from the corresponding generalized distances within each of these two sectors, and averaging over a sufficiently large disorder ensemble, we obtain
\begin{align}
V_{\rm gen} =&\left\langle
\frac{1}{N}\sum_{n,k,m}
|A_{nk}|^2|A_{nm}|^2G(k,m)
\right\rangle_{\rm dis}
\nonumber\\
\approx &
\left\langle
\frac{1}{N}\sum_{n,k}|A_{nk}|^4
\right\rangle_{\rm dis} \times
\left\langle
\frac{1}{N}\sum_k G(k,k)
\right\rangle_{\rm dis}
\nonumber\\
+&\left\langle
\frac{1}{N}\sum_{n}\sum_{k\ne m}
|A_{nk}|^2|A_{nm}|^2
\right\rangle_{\rm dis} \nonumber \\
\times &\left\langle
\frac{1}{N(N-1)}
\sum_{k\ne m}G(k,m)
\right\rangle_{\rm dis}.
\label{eq:separ_aver}
\end{align}

We now define the quantity ${\rm IPR}_T^A$, which measures the
localization of the eigenstates of the complete Hamiltonian in the
Anderson eigenbasis rather than in the original site basis:
\begin{equation}
\langle {\rm IPR}_T^A\rangle
\equiv
\left\langle
\frac{1}{N}\sum_{k,n}|A_{nk}|^4
\right\rangle_{\rm dis}.
\label{eq:ipr_AT}
\end{equation}

Choosing the appropriate $g(i,j)$ in Eq.~(\ref{eq:gen_dist_km}) for each quantity of interest, Eq.~(\ref{eq:separ_aver}) 
yields key approximations connecting these quantities. Specifically, by noting that, at moderate disorder, the wave function overlaps of the different Anderson-localized wave functions are negligible, we find that ${\rm IPR}$ factorizes into a contribution describing delocalization between different Anderson orbitals ($\langle {\rm IPR}^A_T \rangle$) and a contribution describing delocalization within an Anderson orbital ($\langle \mathrm{IPR}_A \rangle$): 
\begin{equation}
\langle {\rm IPR}_T\rangle
\simeq
\langle {\rm IPR}_T^A\rangle
\langle {\rm IPR}_A\rangle.
\label{eq:ipr_T_stat}
\end{equation}
Similar logic leads to the corresponding relation for the spatial variance
\begin{equation}
\langle\sigma_T^2\rangle
\simeq
\langle {\rm IPR}_T^A\rangle
\langle\sigma_A^2\rangle
+
\left(1-\langle {\rm IPR}_T^A\rangle\right)
\sigma_{\rm ex}^2.
\label{eq:sigma_T_stat}
\end{equation}
Detailed derivations of Eqs.~(\ref{eq:ipr_T_stat}) and~(\ref{eq:sigma_T_stat}) are presented in Appendix~\ref{ap1}; see the discussions leading to Eqs.~(\ref{eq:ipr_t_3rd_step}) and~(\ref{eq:sigm_t_3rd_step}), respectively.

Equation~(\ref{eq:sigma_T_stat}) provides a direct interpretation of the
numerical curves: the spatial variance $\langle \sigma^2_T \rangle$ is
interpolated between the Anderson variance $\langle \sigma^2_A \rangle$ and the extended-state variance $\sigma^2_{\rm ex}$ according to the degree of localization of the full eigenstates in the Anderson eigenbasis. In other words, the spatial variance can be viewed approximately as a combination of
the local spatial spread within an Anderson eigenstate, weighted by
$\langle {\rm IPR}^A_T\rangle$, and the long-range spatial spread associated with the mixing of different Anderson eigenstates through the long-range
hopping terms, weighted by
$1-\langle {\rm IPR}^A_T\rangle$.

While the shape of $\langle \sigma^2_A\rangle$ is well established in the literature~\cite{Ozawa2019,Kantelhardt2002,Stone1983}, an explicit analytical expression for $\langle {\rm IPR}^A_T\rangle$ remains unavailable. In the following Sections~\ref{sec_ssv} and~\ref{sec_period}, we analyze different aspects of its behavior across distinct disorder regimes.

\section{Comparison with numerical data}
\label{sec_test_num}

The numerical results related to the analytical predictions presented in Section~\ref{sec_anderson} are shown in Fig.~\ref{fig:comb_ipr_sgm}. The circled yellow dots in panels (a), (b), and (c) show the numerical calculation for the transitional inverse participation ratio ${\rm IPR}^A_T$ introduced in Eq.~(\ref{eq:ipr_AT}) over the wide range of disorder strength $W$ for the three different $\eta_j$ configurations.

For the uniform case, $\eta_j=1$, the results in Fig.~\ref{fig:comb_ipr_sgm}(a) show strong localization of the complete Hamiltonian in the Anderson eigenbasis over a range of weak disorder. The value of ${\rm IPR}^A_T$ remains substantially higher than ${\rm IPR}_R$ applicable to the Richardson limit $J=0$, even in the range of $W$ where the Anderson spatial variance $\langle\sigma^2_A\rangle$ has already become substantially contracted. The randomness introduced into $\eta_j$ significantly reduces this behavior. Figure~\ref{fig:comb_ipr_sgm}(b) shows much weaker localization of the eigenstates in the Anderson eigenbasis for the case $\eta_j\in[0,1]$. Figure~\ref{fig:comb_ipr_sgm}(c) demonstrates partial localization of the eigenstates in the Anderson eigenbasis beginning from very low disorder for the random case $\eta_j\in[-1,1]$. The resulting dependence closely resembles that of ${\rm IPR}_R$ on $W$ in the Richardson limit $J=0$, where ${\rm IPR}_R$ starts from ${\rm IPR}_R\approx0.58$ at low $W$ (for nonuniform $\eta_j$) and remains approximately at this level until $W$ reaches $W_{\rm cr,R}$, beyond which the system becomes completely localized.

The blue lines labeled $\langle {\rm IPR}^A_T \rangle \langle {\rm IPR}_A \rangle$ in Fig.~\ref{fig:comb_ipr_sgm}(a)--(c) closely match the blue dots representing $\langle {\rm IPR}_T \rangle$. This demonstrates excellent agreement between the expression in Eq.~(\ref{eq:ipr_T_stat}) and the numerical results for both uniform and non-uniform $\eta_j$ cases.

The analytical results for $\langle \sigma^2_T \rangle$ obtained using Eq.~(\ref{eq:sigma_T_stat}) are shown by the blue lines labeled $\langle \sigma^2_T({\rm IPR}^A_T) \rangle$ in Fig.~\ref{fig:comb_ipr_sgm}(d)--(f). The corresponding numerical results for $\langle \sigma^2_T \rangle$ are represented by the blue dots.

For the uniform case, $\eta_j=1$, shown in Fig.~\ref{fig:comb_ipr_sgm}(d), Eq.~(\ref{eq:sigma_T_stat}) shows excellent agreement with the numerical dependence of the spatial variance on $W$.

At the same time, Eq.~(\ref{eq:sigma_T_stat}) does not accurately reproduce the behavior of $\langle \sigma^2_T(W) \rangle$ over the very-low-disorder range for the non-uniform $\eta_j$ cases shown in Fig.~\ref{fig:comb_ipr_sgm}(e) and (f). This discrepancy can be attributed to neglecting the interference term in the approximation for $p^T_k(j)$ in Eq.~(\ref{eq:pk_approx}). As expected, the interference term becomes less significant, and Eq.~(\ref{eq:pk_approx}) becomes more reliable, over the range of $W$ where the Anderson eigenstates are already substantially localized. Therefore, Eq.~(\ref{eq:sigma_T_stat}) remains in very good agreement with the numerical $\langle \sigma^2_T(W) \rangle$ results over the range of disorder where the SSV, DET, and DIT regimes are formed, which is the most essential part of our analysis. 

\begin{figure}[t]
\centering
\includegraphics[width=\columnwidth]{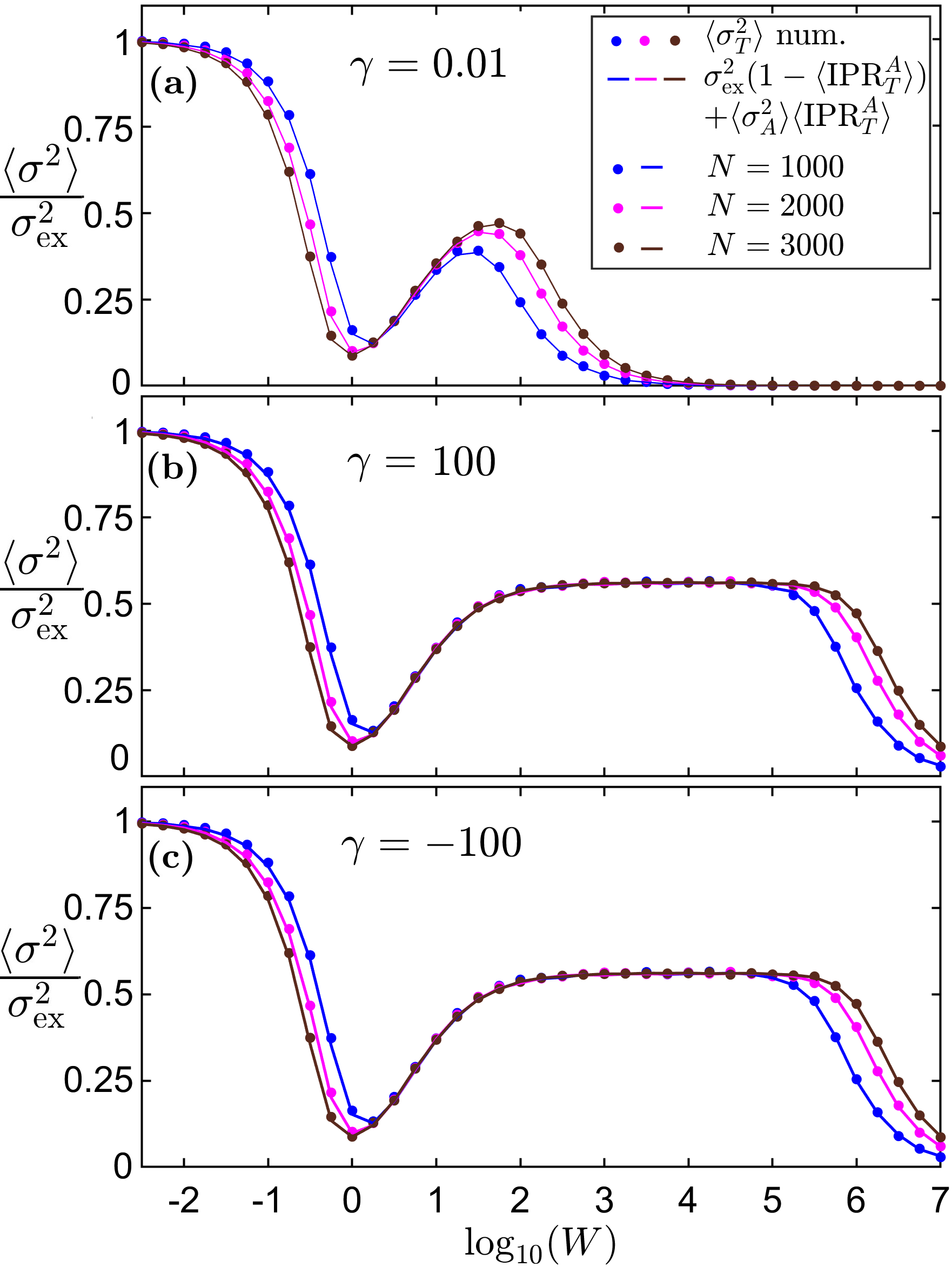}
\caption{System response to disorder over the wide range of disorder strengths $W \in [10^{-2},10^{7}]$, characterized by the $\langle \sigma^2 \rangle$ quantities. The three panels correspond to different choices of the long-range hopping parameters: (a) $\gamma=0.01$; (b) $\gamma = 100$; and (c) $\gamma = -100$. Each panel shows numerical results (colored dots) for the spatial variance of the complete Hamiltonian $H$ for three different system sizes, $N=1000$ (blue), $N=2000$ (magenta), and $N=3000$ (brown), together with the corresponding analytical results (colored lines) from Eq.~\ref{eq:sigma_T_stat} for $\langle \sigma^2_T \rangle$. All numerical calculations are performed for uniform $\eta_j=1$. Averages are taken over all eigenstates and nine disorder realizations. Error bars are smaller than the marker size.
}
\label{fig:parm_samples}
\end{figure}

As seen in~\cite{Chavez2021}, the crossover scales for DET and DIT depend on system size.
Fig.~\ref{fig:parm_samples} demonstrates the behavior of systems with different system sizes and coupling strengths. In all cases, the dependence of the spatial variance on (W) remains in good agreement with the analytical prediction in Eq.~(\ref{eq:sigma_T_stat}). As expected, a larger magnitude of $\gamma$ shifts the $W_{\rm cr}$ threshold to higher disorder strengths, thereby extending the DIT regime, as shown in Figs.~\ref{fig:parm_samples}(b) and (c). In contrast, very small $\gamma$, associated with a smaller $W_{\rm cr}$, shortens the DIT regime to a brief recovery from the SSV regime, as shown in Fig.~\ref{fig:parm_samples}(a). Changing the sign of $\gamma$, as shown in Fig.~\ref{fig:parm_samples}(c), does not noticeably change the key quantities or the localization response to disorder. This behavior is expected. According to~\cite{Modak2016}, changing the sign of $\gamma$ only affects the localization of the ground or highest-energy states that are separated by a gap from the bulk energy spectrum.

\section{Origin of the SSV regime}
\label{sec_ssv}

\begin{figure*}[t]
\centering
\includegraphics[width=\textwidth]{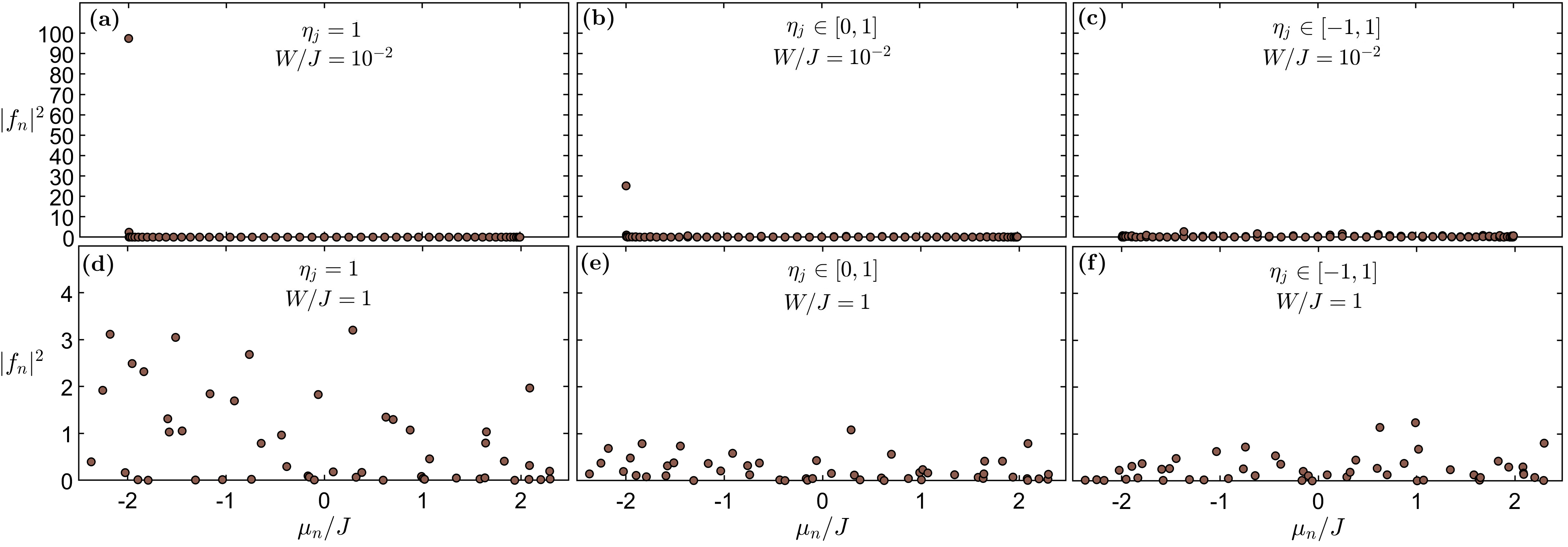}
\caption{$f_n$ coefficients obtained for $N=100$. All numerical results are shown for a single random disorder realization. For uniform $\eta_j=1$, an isolated ground state with a disproportionately large $|f_n|^2$ is observed at low $W$. This effect is also present, although substantially weaker, for $\eta_j\in[0,1]$ and is practically absent for $\eta_j\in[-1,1]$.
}
\label{fig:f2_100_samp}
\end{figure*}

The results discussed in Section.~\ref{sec_test_num} demonstrate that the strong localization of the complete Hamiltonian in the Anderson eigenbasis at weak disorder is the main mechanism underlying
the formation of the SSV regime. This localization can be traced primarily to the uniform long-range amplitudes, $\eta_j=1$, as can be
understood from the $W\to0$ limit. When $W=0$, the eigenstates of the Hamiltonian containing only the nearest-neighbor hopping term are also
eigenstates of both the complete Hamiltonian $H$ and the long-range Hamiltonian $H_L$ for uniform long-range amplitudes $\eta_j=1$.
Consequently, in the absence of disorder, the Anderson eigenbasis
coincides with the eigenbasis of both $H$ and $H_L$. We therefore
expect that, for small but finite $W$, the eigenstates of $H$ remain
strongly localized in the Anderson eigenbasis.

This can also be understood more formally by considering the solution of $H$ in the Anderson eigenbasis, as discussed in Section~\ref{sec_anderson}.

The range of the Anderson eigenvalues can be estimated as $\mu_n \in [-2J-W/2,\,2J+W/2]$~\cite{PhysRevE.100.022142}. However, the average level spacing, $\delta=\mu_{n+1}-\mu_n$, may vary significantly across different parts of the spectrum.

Using the solution of the Richardson model presented in~\cite{Modak2016}, we estimate the participation ratio of a state within a spectral region characterized by an average level spacing $\delta$ as
\begin{align}
\langle PR \rangle \approx \frac{3+3\left(\frac{\delta}{\pi \gamma_{\rm eff}}\right)^2}{1+3\left(\frac{\delta}{\pi \gamma_{\rm eff}}\right)^2}.
\label{eq:est_pr_delta}
\end{align}
Here, $\gamma_{\rm eff}$ accounts for the contribution of remote states to the summation in the secular equation~(\ref{eq:sec_anderson_bas}), as well as the average magnitude of $|f_n|^2$. The participation ratio can therefore range from its maximum value of $\langle PR \rangle \approx 3$ to $\langle PR \rangle = 1$, corresponding to a completely localized state. Thus, for all states to be localized, such that ${\rm IPR}^A_T$ approaches unity, the following condition must be satisfied across the entire spectrum, including regions with the smallest level spacing $\delta$:
\begin{align}
\gamma_{\rm eff} \ll \delta.
\label{eq:cond_ipr_1}
\end{align}

For $W\to0$, the Anderson eigenvalues approach $\mu_n\approx-2J\cos(2\pi n/N)$. Consequently, the level spacing $\delta$ can be as small as $J(2\pi/N)^2$ near the upper and lower edges of the spectrum and as large as $2J(2\pi/N)$ near the center of the band. Within the range
$J(2\pi/N)^2\lesssim W\lesssim 2J(2\pi/N)$,
the smallest level spacing is set by the disorder, $\delta\sim W$. At high disorder strength, $W\gg J$, the level spacing becomes approximately uniform across the spectrum and approaches $\delta\approx W/N$ in the limit $W\to\infty$.

To estimate the $\gamma_{\rm eff}$ side of Eq.~(\ref{eq:cond_ipr_1}), we must examine the values of $f_n$ -- the overlap of the Anderson eigenstate $\psi^A_n(j)$ with the long-range structure factor $\eta_j$ -- defined by Eq.~(\ref{eq:fn}). These values are not uniform among the Anderson eigenstates; instead, certain Anderson eigenstates produce $f_n$ values with significantly larger magnitudes than others.

Figure~\ref{fig:f2_100_samp} demonstrates examples of the $f_n$ coefficients for a small system of size $N=100$. All numerical results are shown for a single random disorder realization. The values of $|f_n|^2$ are shown over the entire spectrum of $\mu_n$ for uniform and non-uniform $\eta_j$ configurations and for different disorder strengths $W$, revealing significantly different distributions of $|f_n|^2$. Each column of panels in Fig.~\ref{fig:f2_100_samp} shows data for the same $\eta_j$ configuration, while each row shows results for a single disorder strength. For the uniform case, an isolated ground state with a disproportionately large $|f_n|^2$ is observed at low $W$. The distribution becomes completely mixed at higher $W$. This effect is also observed, although to a much lesser degree, for the $\eta_j\in[0,1]$ case and is practically absent for $\eta_j\in[-1,1]$.

This behavior can be explained 
by remembering that, as $W\to 0$, the
Anderson eigenstates approach momentum states,
\begin{equation}
\psi_k^A(j)\simeq
\frac{1}{\sqrt N}
\exp\left(\frac{2\pi i k j}{N}\right).
\label{eq:psi_moment}
\end{equation}
For $\eta_j=1$,
\begin{align}
f_k
&=
\sum_j\psi_k^A(j)
\simeq
\frac{1}{\sqrt N}
\sum_j
\exp\left(\frac{2\pi i k j}{N}\right),
\label{eq:moment_sum}
\end{align}
and hence
\begin{equation}
|f_k|^2\simeq
\begin{cases}
N,& k=0,\\
0,& k\ne0.
\end{cases}
\label{eq:fn_N_0}
\end{equation}
Thus, the uniform long-range coupling selectively couples to the symmetric $k=0$ Anderson state through a disproportionately large matrix element.

Consider a root of the secular equation~(\ref{eq:sec_anderson_bas}) located between $\mu_p$ and $\mu_{p+1}$ and away from the ground-state energy. The large $|f_0|^2$ term dominates the secular equation, giving schematically
\begin{equation}
\sum_{n\ne0}
\frac{|f_n|^2}{\lambda-\mu_n}
\simeq
-\frac{|f_0|^2}{\mu_p-\mu_0}
-\frac{1}{\gamma}.
\label{eq:eff_sec}
\end{equation}

This is equivalent to a secular equation with a very large right-hand side and a very small average $|f_{n\ne0}|^2$ on the left-hand side. For low disorder, the energy separation between $\mu_p$ and $\mu_0$ is at least of order $J$, $\mu_p-\mu_0\sim J$. Therefore, the effective parameter $\gamma_{\rm eff}$ entering the secular equation for the excited states in Eq.~(\ref{eq:eff_sec}) can reach a maximum value estimated as
\begin{equation}
(\gamma_{\rm eff})_{\rm max}
\sim \frac{J \langle |f_{n \ne 0}|^2 \rangle}{|f_0|^2}.
\label{eq:gamma_eff}
\end{equation}

A sufficiently small $(\gamma_{\rm eff})_{\rm max}$ can then satisfy the inequality in Eq.~(\ref{eq:cond_ipr_1}) throughout the low-disorder regime, including the limit $W\to0$, even for the smallest possible level spacing $\delta$. This requires
\begin{align}
(\gamma_{\rm eff})_{\rm max} \ll J\left(\frac{2 \pi}{N}\right)^2.
\label{eq:y_eff_req}
\end{align}

\begin{figure*}[t]
\centering
\includegraphics[width=\textwidth]{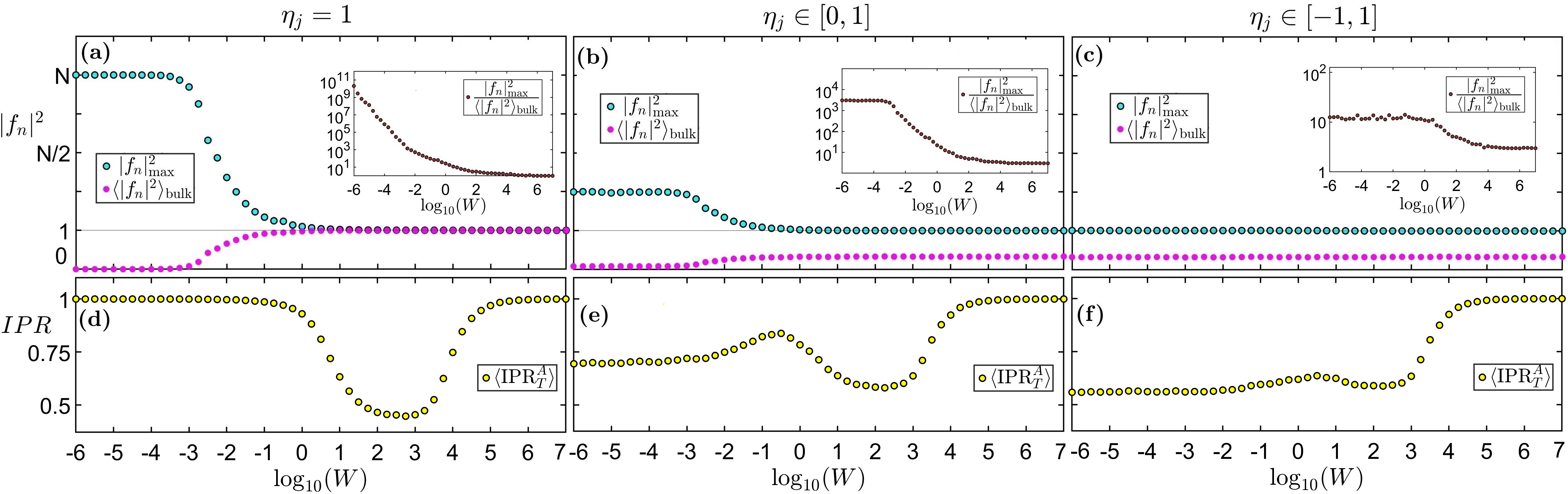}
\caption{System response to disorder over the wide range of disorder strengths $W \in [10^{-6},10^{7}]$, characterized by the $|f_n|^2$ and ${\rm IPR}^A_T$ quantities. 
The first row, panels (a)--(c), shows the statistics of the $|f_n|^2_{\rm max}$ (light blue rounded dots) and $\langle |f_n|^2_{\rm bulk} \rangle$ (rose dots) values used in the Anderson eigenbasis solution discussed in Section~\ref{sec_anderson} and defined by Eq.~(\ref{eq:fn}). Insets in panels (a)--(c) show the evolution of the ratio $|f_n|^2_{\rm max}/\langle |f_n|^2_{\rm bulk} \rangle$, which is the main factor determining $\gamma_{\rm eff}$ in Eq.~(\ref{eq:gamma_eff}). 
The second row, panels (d)--(f), shows ${\rm IPR}^A_T$ (yellow rounded dots), which characterizes the localization of the complete Hamiltonian $H$ in the Anderson eigenbasis. 
All numerical calculations are performed for a system size of $N=1000$ and long-range coupling strength $\gamma=1$. Averages are taken over all eigenstates and nine disorder realizations. Error bars are smaller than the marker size.
}
\label{fig:comb_f2_avr}
\end{figure*}

The detailed evolution of the maximum $|f_n|^2$ value of the system
(associated with $|f_0|^2$ in the above discussion) and the bulk
average of the remaining $|f_n|^2$ values is shown in
Fig.~\ref{fig:comb_f2_avr}(a)--(c). As expected, for $\eta_j=1$ [Fig.~\ref{fig:comb_f2_avr}(a)], the
maximum $|f_n|^2_{\rm max}$ approaches $N$ in the low-$W$ limit and
decreases toward $1$ at high $W$, where each $k$th Anderson eigenstate
becomes localized around its corresponding site $j_k$,
$\psi^A_k(j) \approx \delta_{j,j_k}$. The bulk average starts from
nearly zero at low $W$ and approaches $1$ as the Anderson eigenstates
become localized. The insets of Fig.~\ref{fig:comb_f2_avr}(a)--(c) show
an important quantity,
$|f_n|^2_{\rm max} / \langle |f_n|^2_{\rm bulk} \rangle$, which is used
in the calculation of $\gamma_{\rm eff}$ in Eq.~(\ref{eq:gamma_eff}). In the uniform case [Fig.~\ref{fig:comb_f2_avr}(a)], this
ratio exhibits an unbounded exponential growth as $W$ approaches $0$, readily satisfying the condition in Eq.~(\ref{eq:y_eff_req}).

For the random case shown in Fig.~\ref{fig:comb_f2_avr}(b), with
$\eta_j \in [0,1]$, $|f_n|^2_{\rm max}$ has a lower value in the $W\to0$ limit, associated with the squared average $\langle \eta_j \rangle^2 \approx 1/4$ multiplied by $N$. More importantly, however, $\langle |f_n|^2_{\rm bulk} \rangle$ remains finite and nonzero even at very low $W$, which significantly limits the ratio $|f_n|^2_{\rm max} / \langle |f_n|^2_{\rm bulk} \rangle$, as shown in
the inset of Fig.~\ref{fig:comb_f2_avr}(b), thereby resulting in a larger
$\gamma_{\rm eff}$. In contrast to the
uniform $\eta_j=1$ and random $\eta_j \in [0,1]$ cases, the results for the random $\eta_j \in [-1,1]$ system in Fig.~\ref{fig:comb_f2_avr}(c) show only occasional $|f_n|^2$ values that happen to exceed the bulk average. Also note that, for nonuniform $\eta_j$, $\langle |f_n|^2_{\rm bulk} \rangle$ reaches a maximum below unity, consistent with the averaged squared amplitude $\langle |\eta_j|^2 \rangle \approx 1/3$. 

For a side-by-side comparison, Fig.~\ref{fig:comb_f2_avr}(d)--(f) are included to show the evolution of ${\rm IPR}^A_T$, previously presented in Fig.~\ref{fig:comb_ipr_sgm}(a)--(c). As expected, ${\rm IPR}^A_T$ falls below unity at approximately the same disorder strength at which $|f_n|^2_{\rm max}$ reaches its minimum value of $1$.

Overall, we conclude that, for uniform $\eta_j$, the resulting transformation coefficients $A_{nk}$ become concentrated around individual Anderson states, such that
\begin{equation}
\langle {\rm IPR}^A_T\rangle\to 1
\end{equation}
in the low-disorder regime.
Eq.~(\ref{eq:sigma_T_stat}) then yields a small spatial variance, which accounts for the dip observed in
Fig.~\ref{fig:comb_ipr_sgm}(d).

For random $\eta_j$, the overlap
\begin{equation}
f_n=\sum_j\eta_j\psi_n^A(j)
\end{equation}
no longer isolates the symmetric state in the same way. The localization in the Anderson eigenbasis is therefore weakened already at small $W$, as seen directly in Figs.~\ref{fig:comb_ipr_sgm}(e) and (f).
Through Eq.~(\ref{eq:sigma_T_stat}), this produces a larger spatial variance and largely removes the low-disorder variance minimum associated with the SSV regime.

\section{Periodic long range amplitude.}
\label{sec_period}

\begin{figure*}[t]
\centering
\includegraphics[width=\textwidth]{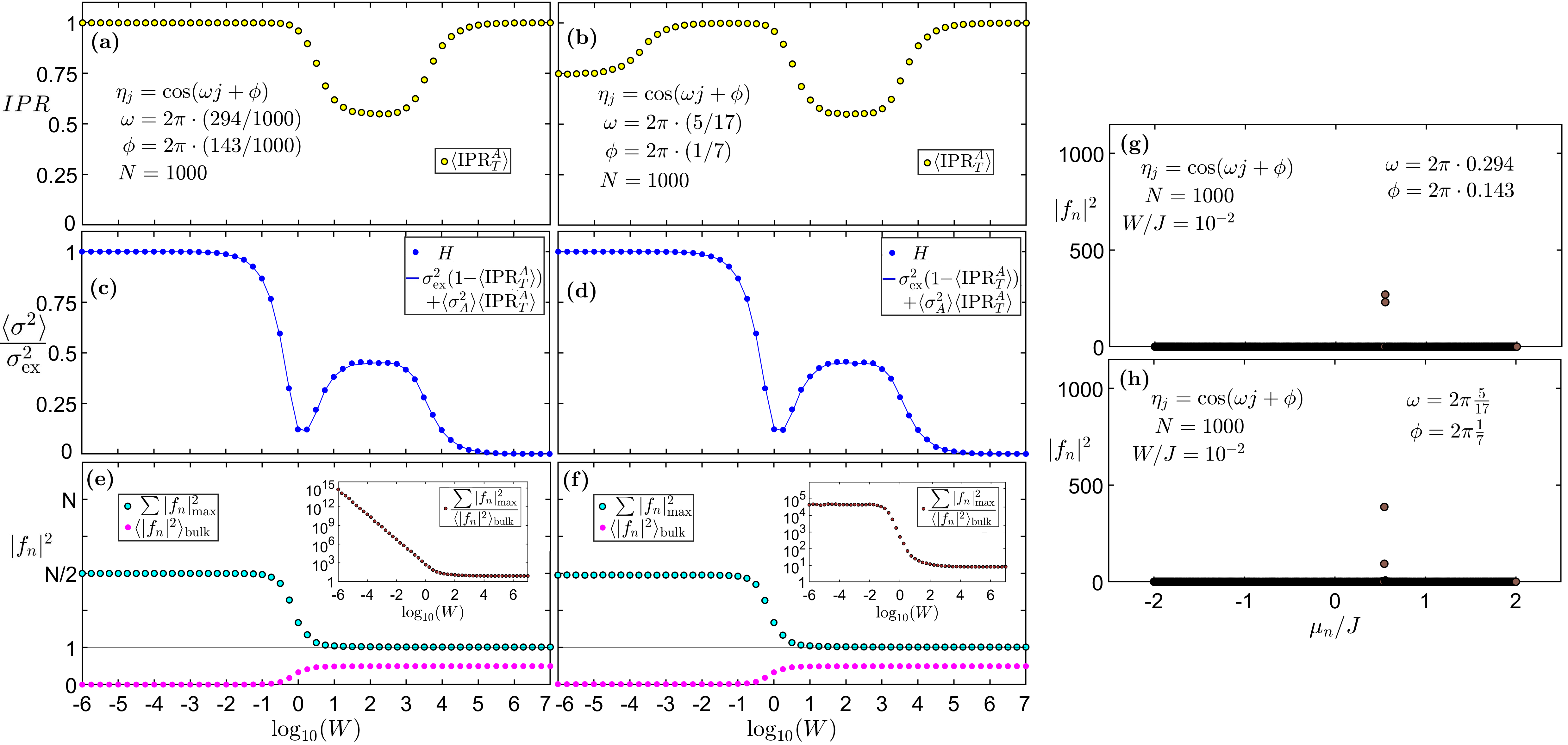}
\caption{System response to disorder over the wide range of disorder strengths $W \in [10^{-6},10^{7}]$ in the presence of periodic long-range amplitudes $\eta_j=\cos(\omega j+\phi)$. Calculations are performed for two nearly identical sets of parameters. Panels (a), (c), (e), and (g) correspond to $\omega = 2\pi\cdot0.294$ and $\phi = 2\pi\cdot0.143$, for which $\omega/(2\pi)=294/1000$, exactly commensurate with the finite system size $N=1000$.
Panels (b), (d), (f), and (h) correspond to $\omega = 2\pi\cdot(5/17)$ and $\phi = 2\pi\cdot(1/7)$. Panels (a) and (b) show the inverse participation ratio ${\rm IPR}^A_T$ of the $H$ eigenstates in the Anderson eigenbasis. Panels (c) and (d) show the spatial variance of the complete Hamiltonian $H$. Panels (e) and (f) show the statistics of $\sum_{k \in {\rm max}}|f_k|^2$ (labeled as $\sum |f_k|^2_{\rm max}$) and $\langle |f_{k \in {\rm bulk}}|^2 \rangle$ (labeled as $\langle|f_k|^2\rangle_{\rm bulk}$), used in the Anderson eigenbasis solution discussed in Sec.~\ref{sec_anderson} and defined by Eq.~(\ref{eq:fn}). Panels (g) and (h) show the actual $|f_k|^2$ values across the entire $\mu_k$ spectrum, calculated at disorder strength $W/J = 10^{-2}$ for a single random disorder realization. All numerical calculations are performed for a system size of $N=1000$ and long-range coupling strength $\gamma=1$. Averages in panels (a)--(f) are taken over all eigenstates and nine disorder realizations. Error bars are smaller than the marker size.
}
\label{fig:period}
\end{figure*}

An important observation follows from the discussion in
Section~\ref{sec_ssv} concerning the origin of the SSV regime. In
particular, eliminating this regime requires sufficient randomness in
the $\eta_j$ parameters. To illustrate this point, consider a periodic
form,
$\eta_j=\cos(\omega j+\phi)$, which may appear natural given the
standing-wave cavity-mode profile shown in Fig.~\ref{fig:model}. In this
case, the sum in Eq.~\ref{eq:moment_sum} takes the form
\begin{align}
f_k
&=
\sum_j \eta_j \psi_k^A(j)
\nonumber \\
&\simeq
\frac{1}{2\sqrt N}
\sum_j \exp\left(i \left(\frac{2\pi k }{N} + \omega \right) j + \phi\right) \nonumber \\
&+ \frac{1}{2\sqrt N} \sum_j
\exp\left(i \left(\frac{2\pi k}{N} - \omega \right) j- \phi\right),
\label{eq:moment_period_sum}
\end{align}
which implies that, for $N\gg1$, there will generally be two to four
values of $k$ for which $2\pi k/N\pm\omega$ is close to an integer
multiple of $2\pi$. The corresponding sums then produce
disproportionately large values of $|f_k|^2$, while the remaining
$|f_k|^2$ values remain close to zero.

Thus, a periodic structure in $\eta_j$ also isolates specific Anderson states, in contrast to a genuinely random spatial profile. In the $W\to0$ limit, these states have energies near $E=-2 J \cos(\omega)$, analogous to the symmetric ground state isolated by the uniform profile $\eta_j=1$.

Following the same logic used in Sec.~\ref{sec_ssv} to derive $\gamma_{\rm eff}$, Eq.~(\ref{eq:gamma_eff}) can now be generalized to include all disproportionately large $|f_k|^2$ values, denoted by $|f_{k\in{\rm max}}|^2$, whose corresponding Anderson eigenvalues $\mu_k$ are concentrated within a narrow energy interval around $E=-2J\cos(\omega)$:
\begin{equation}
\gamma_{\rm eff}
= \frac{J \langle |f_{k \in {\rm bulk}}|^2 \rangle}{\sum_{k \in {\rm max}} |f_k|^2} .
\label{eq:gamma_eff_sum}
\end{equation}
where $\langle |f_{k \in {\rm bulk}}|^2 \rangle$ denotes the average over the remaining $|f_k|^2$ values that do not belong to this isolated group.

Depending on $\omega$, the exact equality
$2\pi k/N\pm\omega=2\pi n$ may be satisfied for certain values of $k$.
In this case, $\gamma_{\rm eff}\to0$ as $W\to0$, which in turn leads to
complete localization of the $H$ eigenstates in the Anderson
eigenbasis already at very low disorder. If the closest approximation
$2\pi k/N\pm\omega\approx2\pi n$ is not exact, $\gamma_{\rm eff}$ remains
finite, and ${\rm IPR}^A_T$ reaches $1$ only above the critical disorder
scale at which the level spacing becomes $\delta\sim W$, such that
$\gamma_{\rm eff} \ll \delta$.

To demonstrate the effect of periodicity in $\eta_j$, we perform numerical calculations for two different values of $\omega$, with the results shown in Fig.~\ref{fig:period}: $\omega = 2\pi\cdot(5/17)$ and $\omega = 2\pi\cdot0.294$. Although these two values are equal up to rounding error, only the latter can satisfy the condition $2\pi k/N\pm\omega = 2 \pi n$ for $N = 1000$ and integer $k$ and $n$, which leads to a significant change in the behavior of ${\rm IPR}^A_T(W)$ as $W\to0$.

The evolutions of $\langle \sigma^2_T(W) \rangle$ for both values of $\omega$, shown by the blue dots in Fig.~\ref{fig:period}(c),(d), are nearly identical and closely reproduce the dependence of $\langle \sigma^2_T \rangle$ on $W$ observed for the uniform case $\eta_j=1$. The main difference is a lower DIT plateau, resulting from the lower $\langle \sigma^2_R \rangle$ associated with non-uniform $\eta_j$, as discussed in Section~\ref{sec_richardson}. In both cases, $\omega=2\pi\cdot(5/17)$ and $\omega=2\pi\cdot0.294$, $\langle \sigma^2_T(W) \rangle$ closely follows the analytical expression in Eq.~\ref{eq:sigma_T_stat}, shown by the blue line.

Although this small difference in $\omega$ does not affect the final behavior of $\langle \sigma^2_T(W) \rangle$, it qualitatively changes the localization of the $H$ eigenstates in the Anderson eigenbasis as $W\to0$, as shown in Fig.~\ref{fig:period}(a),(b). As discussed above, $\langle {\rm IPR}^A_T\rangle\to1$ as $W\to0$ for $\omega=2\pi\cdot0.294$ [Fig.~\ref{fig:period}(a)], whereas $\langle {\rm IPR}^A_T\rangle$ remains below $1$ over a range of very small $W$ values for $\omega=2\pi\cdot(5/17)$ [Fig.~\ref{fig:period}(b)]. This pronounced sensitivity to a small change in the periodicity of the long-range amplitudes further supports the proposed localization mechanism and the corresponding estimate of $\gamma_{\rm eff}$.

Fig.~\ref{fig:period}(e),(f) show the evolution of the $|f_k|^2$ coefficients over the range of disorder strengths. While the sum of the disproportionately large $|f_k|^2$ values is nearly identical for the two $\omega$ cases, the average of the remaining $|f_k|^2$ values differs significantly. The inset in Fig.~\ref{fig:period}(e) shows that $J/\gamma_{\rm eff}$, labeled as $\sum|f_n|^2_{\rm max}/\langle|f_n|^2\rangle_{\rm bulk}$, exhibits unbounded growth as $W\to0$ for $\omega=2\pi\cdot0.294$, resulting in ${\rm IPR}^A_T=1$ throughout the low-disorder region. In contrast, $J/\gamma_{\rm eff}$ remains finite as $W\to0$ for $\omega=2\pi\cdot(5/17)$ [inset in Fig.~\ref{fig:period}(f)], thereby failing to satisfy the inequality in Eq.~(\ref{eq:cond_ipr_1}) for part of the states and causing $\langle {\rm IPR}^A_T \rangle$ to drop below $1$ at low $W$. However, as $W$ increases to a finite value such that Eq.~(\ref{eq:cond_ipr_1}) is satisfied with the level spacing $\delta\sim W$, the system becomes fully localized in the Anderson eigenbasis and enters the SSV regime, with $\langle {\rm IPR}^A_T \rangle = 1$.

Finally, Fig.~\ref{fig:period}(g),(h) show the actual $|f_k|^2$ values across the entire $\mu_k$ spectrum, calculated at disorder strength $W/J=10^{-2}$ for a single random disorder realization. Consistent with the above analysis, isolated, exceptionally large $|f_k|^2$ values appear around $\mu_k\approx-2J\cos(\omega)\approx0.55$.

\section{Effective Current}
\label{sec_current}

\begin{figure*}[t]
\centering
\includegraphics[width=\textwidth]{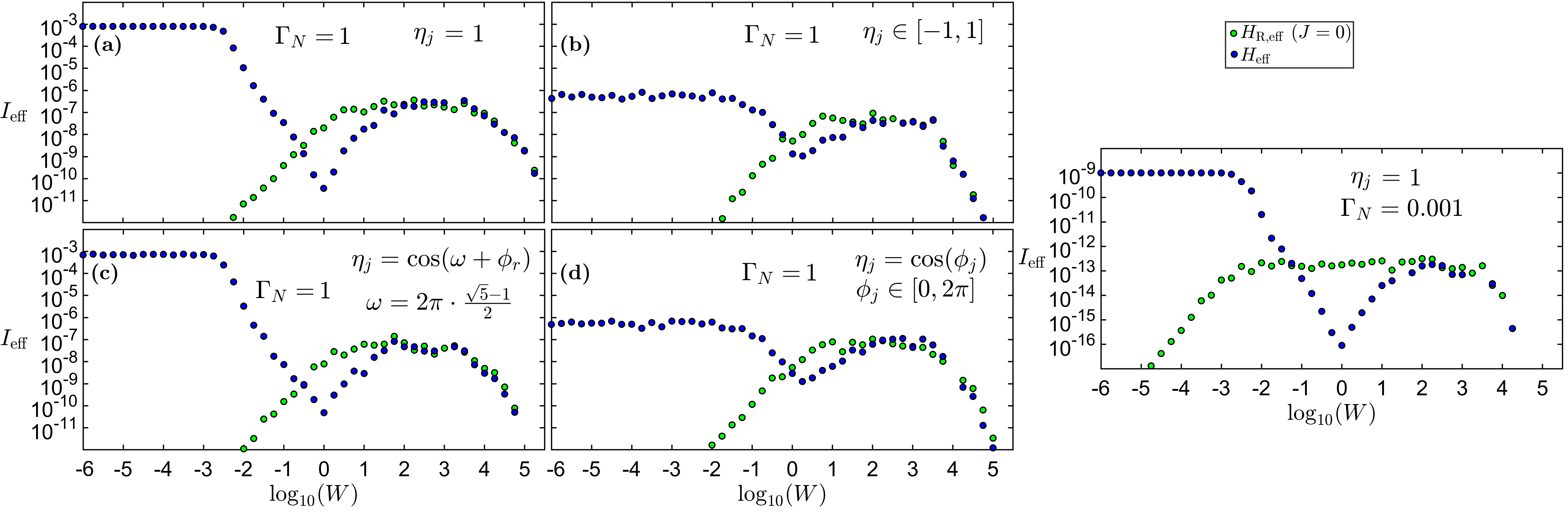}
\caption{The response of the effective current $I_{\rm eff}$ to disorder over the wide range of disorder strengths $W \in [10^{-6},10^{6}]$ for different configurations of the long-range hopping amplitudes $\eta_j$. Panels (a)--(e) correspond to (a) and (e) uniform amplitudes, $\eta_j=1$; (b) randomized amplitudes, $\eta_j\in[-1,1]$; (c) periodic amplitudes, $\eta_j=\cos(\omega j+\phi_r)$, with $\omega=2\pi \cdot (\sqrt{5} - 1) / 2$ and $\phi_r$ - random phase chosen once for each disorder realization; and (d) amplitudes determined by random phases, $\eta_j=\cos(\phi_j)$, with $\phi_j\in[0,2\pi]$. 
Panels (a)--(d) correspond to $\Gamma_N = 1$, panel (e) to $\Gamma_N = 0.001$.
All numerical calculations are performed for a system governed by Eq.~\ref{eq:Lindblad}, with system size $N=1000$ and long-range hopping strength $\gamma=1$. Averages of $I_{\rm eff}$ are taken over 100 disorder realizations.
}
\label{fig:eff_current_tr}
\end{figure*}

Finally, as in~\cite{Chavez2021}, we show that the localization metrics $\sigma^2$ and $\mathrm{IPR}$ are reliable indicators of transport by calculating the effective current $I_{\rm eff}$ through an open one-dimensional chain. We consider the Lindblad master equation~\cite{MasterLinblad}
\begin{equation}
\frac{d\rho}{dt} = -i[H_{1}, \rho] + \sum_k \left( L_k \rho L_k^\dagger - \frac{1}{2} \{L_k^\dagger L_k, \rho\} \right)
\label{eq:Lindblad}
\end{equation}
describing the dynamics of the model with two dissipators: a pump on the first site,
\begin{align}
L_1 = L_{\text{pump}} = \sqrt{\Gamma_1} c_1^\dagger
\end{align}
and a drain from the last site,
\begin{align}
L_2 = L_{\text{drain}} &= \sqrt{\Gamma_N} c_N.
\end{align}
The Hamiltonian $H_1$ is similar to the Hamiltonian $H$ of the closed system, except that the first and last sites are spatially separated and are not connected by the nearest-neighbor hopping term:
\begin{align}
H = H_1 -J(|N\rangle \langle 1| + |1\rangle \langle N|).
\end{align}

For this qualitative analysis, we avoid the computationally expensive time evolution governed by Eq.~\ref{eq:Lindblad}. Instead, following the transient single-excitation approach employed in~\cite{Kropf2019,Chavez2021},
we estimate the maximum effective current through the system as
\begin{equation}
I_{\rm eff} = \exp\left(\left\langle \ln\left(\frac{1}{\tau}\right) \right\rangle_{\rm dis} \right),
\label{eq:eff_current_tr}
\end{equation}
where $\tau$ is the average transfer time~\cite{Rebentrost2009,Mohseni2008}, defined as
\begin{equation}
\tau = \Gamma_N \int^\infty_0 t |\Psi(t,N)|^2\,{\rm d}t,
\label{eq:tau_def}
\end{equation}
and $\Psi(t,N)$ is the probability amplitude at the drain site $N$ at time $t$, given that a single particle is initially injected at site $1$. The time evolution is governed by the effective non-Hermitian Hamiltonian
\begin{equation}
H_{\rm eff} = H_1 - i\frac{1}{2}\Gamma_N |N\rangle\langle N|,
\end{equation}
where the second term accounts for particle loss through the drain.

The choice of a logarithmic scale in Eq.~\ref{eq:eff_current_tr} is justified by the following statistical argument. When an excitation is initially placed at the first site at $t=0$, its probability distribution spreads across the system over time. The effective current associated with transferring the excitation to a remote drain site at a distance $L$ is therefore expected to depend on the time-averaged probability of reaching that site. Approximating the probability distribution by a Gaussian centered at the initial site, with variance $\sigma^2$, we estimate the probability at distance $L$ as
\begin{align}
p(L) \sim \frac{1}{\sqrt{2 \pi \sigma^2 }}\exp\left(-\frac{L^2}{2 \sigma^2}\right).
\end{align}
Consequently, even a modest change in $\sigma^2$ can lead to an exponential change in the effective current.

The numerical results for the effective current, calculated using Eqs.~(\ref{eq:eff_current_tr}) and~(\ref{eq:tau_def}), are shown in Fig.~\ref{fig:eff_current_tr}. For comparison, we also consider the $J=0$ limit of $H_{\rm eff}$, denoted by $H_{\rm R,eff}$, which corresponds to the Richardson model.

The single-excitation approach captures the strong destructive interference characteristic of the Richardson model at low disorder. In the disorder-free limit, $W\to0$, the time evolution of the site probability amplitudes in the closed Richardson system ($\Gamma_N=0$), with the excitation initially localized at site $j=1$ at time $t=0$, can be obtained exactly:
\begin{align}
\Psi(t, j)=\delta_{j,1}
+\frac{\eta_j \eta_1}{S}
\left[ e^{i \gamma t S}-1 \right]
\end{align}
where $S=\sum_{j=1}^{N}\eta_j^2$.

This expression shows that only a small fraction of the probability reaches the drain site $j=N$, where it decays through the non-Hermitian term $- i\frac{1}{2}\Gamma_N |N\rangle\langle N|$. Consequently, the effective current remains very small below a certain disorder strength, with the crossover determined by $\Gamma_N$.

The overall dependence on the disorder strength is qualitatively similar to that of $\langle \sigma^2_T(W) \rangle$ discussed in the previous sections with the caveat that for $\Gamma_N = 1$, the effective current for $J = 0$ limit experience its own recovery from the destructive interference over the same disorder range where complete system demonstrates suppression of current associated with SSV regime. 

For the uniform case, $\eta_j=1$, shown in Fig.~\ref{fig:eff_current_tr}(a), the effective current decreases over the small-to-moderate disorder range, followed by its recovery in the disorder-enhanced transport (DET) regime and a plateau corresponding to the disorder-independent transport (DIT) regime, before the current is eventually collapsed above the critical disorder threshold.

Randomizing the long-range amplitudes, $\eta_j\in[-1,1]$, eliminates a substantial portion of the initial decline in the effective current [Fig.~\ref{fig:eff_current_tr}(b)], bringing it closer to the Richardson-model $I_{\rm eff}$ curve. In accordance with the discussion in Sec.~\ref{sec_period}, introducing periodic long-range amplitudes of the form $\eta_j=\cos(\omega j+\phi_r)$, as expected for a standing-wave cavity mode, does not improve transport in the region of pronounced current suppression. The results for the periodic profile shown in Fig.~\ref{fig:eff_current_tr}(c) are obtained using the golden-ratio frequency $\omega = 2\pi(\sqrt{5}-1)/2$, a widely used choice for quasiperiodic systems owing to its strong incommensurability~\cite{Doggen2019,Liebchen2014,Liu2025,AubryAndre1980,Macia2014}. Nevertheless, this incommensuration shows no notable differences from commensurate wavelengths.
The phase $\phi_r$ is chosen randomly and independently for each disorder realization to eliminate the dependence on the specific values of $\eta_1$ and $\eta_N$.

An alternative approach is to make the emitter positions [Fig.~\ref{fig:model}] less periodic, such that the long-range amplitudes are determined by random phases, $\eta_j=\cos(\phi_j)$, with $\phi_j\in[0,2\pi]$, rather than by the periodic phase $\eta_j=\cos(\omega j)$. The results in Fig.~\ref{fig:eff_current_tr}(d) support this approach: the region of current suppression is largely eliminated, similarly to the case of fully random $\eta_j\in[-1,1]$. At the same time, such spatial randomness would likely introduce disorder in the nearest-neighbor hopping amplitudes, which remains a subject for future investigation.

For comparison, an additional calculation for uniform $\eta_j = 1$ and $\Gamma_N = 0.001$ is shown in Fig.~\ref{fig:eff_current_tr}(e). As expected, it exhibits a lower effective current, but also demonstrates delocalization in the Richardson model at a lower disorder strength.

\section{Conclusions and Outlook}
\label{sec_outlook}

We have studied a disordered one-dimensional closed chain with a site-dependent long-range hopping term. The spatial variance defined in Eq.~(\ref{eq:sigma}) provides a natural measure of spatial spreading on a closed topology, for which we obtained the principal analytical result
\begin{equation}
\langle\sigma^2_T\rangle
\simeq
\langle {\rm IPR}^A_T\rangle\langle\sigma_A^2\rangle
+
(1-\langle {\rm IPR}^A_T\rangle)\sigma_{\rm ex}^2.
\end{equation}
This result follows from expanding the full eigenstates in the Anderson basis and averaging over disorder realizations. The quantity $\langle {\rm IPR}^A_T\rangle$ controls the balance between the Anderson variance $\langle\sigma_A^2\rangle$ and the variance of a completely extended state, $\sigma_{\rm ex}^2$, which contribute to the resulting $\langle\sigma^2_T\rangle$.

The large value of $\langle {\rm IPR}^A_T\rangle$ in the low-disorder regime is the main cause of the suppressed spatial variance (SSV) regime. This effect is strongest for uniform all-to-all long-range interactions, for which $\langle {\rm IPR}^A_T\rangle\to1$.
We interpret the disorder-enhanced transport (DET) regime as a recovery from SSV.

At the same time, the exact analytical form of the ${\rm IPR}^A_T(W)$ dependence remains to be established. Such a result would allow more precise predictions of transport behavior across different disorder regimes.

Introducing randomness into the long-range amplitudes $\eta_j$ largely reduces or eliminates the SSV regime, thereby enhancing spatial spreading over the weak-to-moderate disorder range. However, we have shown that this effect cannot be achieved by simple periodic amplitudes of the form $\eta_j=\cos(\omega j+\phi)$, which are expected for a standing-wave cavity mode.
A practical implementation will require randomizing the emitter positions within the cavity, which, in turn, would likely introduce disorder in the nearest-neighbor hopping amplitudes. Investigating experimental configurations and designing devices capable of realizing such spatial randomness therefore represents an important priority for future research.

Another possible direction for future study is to extend this approach to more general models in which the dominance of short-distance interactions may lead to the formation of an SSV regime, in which a particle or excitation is more likely to propagate through a shorter pathway than to reach more distant sites through a long-distance channel. One possible model of interest is described in~\cite{mqw1-jfvb}, which includes power-law distance-dependent hopping amplitudes.

Future research should also investigate many-body scenarios involving multiple particles or excitations propagating through the chain.

\begin{acknowledgments}
This work was performed with support from the National Science Foundation through Awards No. DMR-1945529 and No. MPS-2228725 (M.K.).
\end{acknowledgments}

\onecolumngrid
\appendix

\section{Detailed algebra for the spatial variance and ${\rm IPR}$ calculation}\label{ap1}

We now rewrite the previously introduced quantities by making the
averaging over eigenstates and disorder realizations explicit:
\begin{align}
\langle {\rm IPR}_T\rangle
&\equiv
\left\langle
\frac{1}{N}\sum_{k,j}[p_k^T(j)]^2
\right\rangle_{\rm dis},
\label{eq:ipr_T}\\
\langle {\rm IPR}_A\rangle
&\equiv
\left\langle
\frac{1}{N}\sum_{k,j}[p_k^A(j)]^2
\right\rangle_{\rm dis},
\label{eq:ipr_A}\\
\langle {\rm IPR}_R\rangle
&\equiv
\left\langle
\frac{1}{N}\sum_{k,j}[p_k^R(j)]^2
\right\rangle_{\rm dis},
\label{eq:ipr_R}\\
\langle \sigma_T^2\rangle
&\equiv
\left\langle
\frac{1}{2N}\sum_{k,i,j}
p_k^T(i)p_k^T(j)\,
\operatorname{dist}^2(i,j)
\right\rangle_{\rm dis},
\label{eq:sigma_T}\\
\langle \sigma_A^2\rangle
&\equiv
\left\langle
\frac{1}{2N}\sum_{k,i,j}
p_k^A(i)p_k^A(j)\,
\operatorname{dist}^2(i,j)
\right\rangle_{\rm dis},
\label{eq:sigma_A}\\
\langle \sigma_R^2\rangle
&\equiv
\left\langle
\frac{1}{2N}\sum_{k,i,j}
p_k^R(i)p_k^R(j)\,
\operatorname{dist}^2(i,j)
\right\rangle_{\rm dis}.
\label{eq:sigma_R}
\end{align}

Using the definitions in 
Eqs.~(\ref{eq:ipr_T}-\ref{eq:sigma_R}) and~(\ref{eq:ipr_AT}), together with completeness and normalization identities, we derive a few useful
relations that are convenient for this analysis. The following relations
can be written for any eigenbasis $|\psi_n\rangle$ and its corresponding
site probabilities $p_n(j)\equiv|\psi_n(j)|^2$. They therefore apply to
the quantities $p^{T}(j)$, $p^{A}(j)$, and $p^{R}(j)$:
\begin{align}
    1=&\Big(\sum_{j}p_{n}(j)\Big)^{2}=\sum_{j}p_{n}^{2}(j)+\sum_{j\ne i}p_{n}(j)p_{n}(i)\implies\sum_{j\ne i}p_{n}(j)p_{n}(i)=1-\sum_{j}p_{n}^{2}(j),
    \label{eq:normal_p_j} \\
    1=&\Big(\sum_{n}p_{n}(j)\Big)^{2}=\sum_{n}p_{n}^{2}(j)+\sum_{n\ne k}p_{n}(j)p_{k}(j)\implies\sum_{n\ne k}p_{n}(j)p_{k}(j)=1-\sum_{n}p_{n}^{2}(j).
    \label{eq:normal_p_n}
\end{align}
Similarly, we can write the following relations for the coefficients
$A_{nk}$:
\begin{align}
\sum_{k\ne m}|A_{nk}|^{2}|A_{nm}|^{2}=1-\sum_{k}|A_{nk}|^{4}\implies\sum_{n,k\ne m}|A_{nk}|^{2}|A_{nm}|^{2}=N-\sum_{n,k}|A_{nk}|^{4}=N(1-\langle {\rm IPR}_{T}^{A}\rangle),
\label{eq:normal_a_j} \\
\sum_{k\ne m}|A_{kn}|^{2}|A_{mn}|^{2}=1-\sum_{k}|A_{kn}|^{4}\implies\sum_{k\ne m,n}|A_{kn}|^{2}|A_{mn}|^{2}=N-\sum_{k,n}|A_{kn}|^{4}=N(1-\langle {\rm IPR}_{T}^{A}\rangle).
\label{eq:normal_a_n}
\end{align}

We now calculate ${\rm IPR}_T$ for the complete Hamiltonian:
\begin{equation}
    \langle {\rm IPR}_{T}\rangle\equiv
    \left\langle\frac{1}{N}\sum_{n,j}p_{n}^{T}(j)^{2}
    \right\rangle_{{\rm dis}}\approx
    \left\langle\frac{1}{N}\sum_{n,j}\Big(\sum_{k}|A_{nk}|^{2}p_{k}^{A}(j)\Big)^{2}
    \right\rangle_{{\rm dis}}=
    \left\langle\frac{1}{N}\sum_{n,j,k,m}|A_{nk}|^{2}|A_{nm}|^{2}p_{k}^{A}(j)p_{m}^{A}(j)
    \right\rangle_{{\rm dis}}
    \label{eq:ipr_t_1st_step}
\end{equation}
When $g(i,j)$ in Eq.~(\ref{eq:gen_dist_km}) is chosen as the Kronecker
delta,
\begin{equation}
g(i,j)=\delta_{ij},
\end{equation}
Eqs.~(\ref{eq:ipr_t_1st_step}) and (\ref{eq:separ_aver}), together
with the identities in Eqs.~(\ref{eq:normal_p_j})--(\ref{eq:normal_a_n}),
give the following expression for ${\rm IPR}_T$:
\begin{equation}
{\rm IPR}_{T} \approx \langle {\rm IPR}_{T}^{A}\rangle\langle {\rm IPR}_{A}\rangle+\frac{(1-\langle {\rm IPR}_{T}^{A}\rangle)(1-\langle {\rm IPR}_{A}\rangle)}{N-1}
\label{eq:ipr_t_2st_step}
\end{equation}
For $N \gg 1$, this becomes
\begin{equation}
{\rm IPR}_{T} \approx\langle
{\rm IPR}_{T}^{A}\rangle\langle {\rm IPR}_{A}\rangle
\label{eq:ipr_t_3rd_step}
\end{equation}

When one half of the squared shortest-path distance is substituted for $g(j,i)$ in
Eq.~(\ref{eq:gen_dist_km}),
\begin{equation}
g(i,j)=\frac{1}{2}\operatorname{dist}^2(i,j),
\end{equation}
Eq.~(\ref{eq:separ_aver}) provides a pathway to calculate
$\langle \sigma^2_T\rangle$:
\begin{align}
\langle\sigma_{T}^{2}\rangle
&\equiv\left\langle\frac{1}{2N}\sum_{n,j,i}p_{n}(j)p_{n}(i)\operatorname{dist}^2(j,i)
\right\rangle_{{\rm dis}} \nonumber \\
&\approx\left\langle\frac{1}{2N}\sum_{n,j,i}\Big(\sum_{k}|A_{nk}|^{2}p_{k}^{A}(j)\Big)\Big(\sum_{m}|A_{nm}|^{2}p_{m}^{A}(i)\Big)\operatorname{dist}^{2}(j,i)
\right\rangle_{{\rm dis}}
\nonumber \\
&=\left\langle\frac{1}{2N}\sum_{n,k,m,j,i}|A_{nk}|^{2}|A_{nm}|^{2}p_{k}^{A}(j)p_{m}^{A}(i)\operatorname{dist}^{2}(j,i)
\right\rangle_{{\rm dis}} \nonumber \\
&\approx\left\langle\frac{1}{N}\sum_{n,k}|A_{nk}|^{4}\rangle_{{\rm dis}}\langle\frac{1}{2N}\sum_{k,j,i}p_{k}^{A}(j)p_{k}^{A}(i)\operatorname{dist}^{2}(j,i)
\right\rangle_{{\rm dis}} \nonumber \\
&+\left\langle\frac{1}{N}
\sum_{n, k \ne m}
|A_{nk}|^{2}|A_{nm}|^{2}
\right\rangle_{{\rm dis}}
\cdot
\left\langle\frac{1}{2N(N-1)}\sum_{k \ne m,j,i}p_{k}^{A}(j)p_{m}^{A}(i)\operatorname{dist}^{2}(j,i)
\right\rangle_{{\rm dis}}
\label{eq:sigm_t_1st_step}
\end{align}

The completeness property of $p^A_k(j)$ and the definition of
$\operatorname{dist}_{\rm av}^2$ help us derive the relation for the
off-diagonal contribution in the last term of Eq.~(\ref{eq:sigm_t_1st_step}). Keeping in mind that
$\sigma^2_{\rm ex}=\operatorname{dist}_{\rm av}^2/2$, we write:
\begin{align}
\operatorname{dist}_{\rm av}^2&\equiv
\frac{1}{N^{2}}\sum_{j,i}\operatorname{dist}^{2}(j,i)=\sum_{j,i}\Big(\frac{1}{N}\sum_{k}p^A_{k}(j)\Big)\Big(\frac{1}{N}\sum_{m}p^A_{m}(i)\Big)\operatorname{dist}^2(j,i) \nonumber \\
&= \frac{1}{N^{2}}\sum_{k,j,i}p^A_{k}(j)p^A_{k}(i)\operatorname{dist}^{2}(j,i)
+\frac{1}{N^{2}}
\sum_{k \ne m,j,i}p^A_{k}(j)p^A_{m}(i)\operatorname{dist}^{2}(j,i) \nonumber \\
&\implies \langle\frac{1}{2N^{2}}
\sum_{k \ne m,j,i}p^A_{k}(j)p^A_{m}(i)\operatorname{dist}^{2}(j,i)\rangle_{\rm dis} = \sigma^2_{\rm ex} - \frac{1}{N}\langle\sigma_A^{2}\rangle
\label{eq:off_diag_p}
\end{align}

Then, Eq.~(\ref{eq:sigm_t_1st_step}), together with the identities in
Eqs.~(\ref{eq:normal_p_j})--(\ref{eq:normal_a_n}),(\ref{eq:off_diag_p}), yields
\begin{equation}
    \langle\sigma_{T}^{2}\rangle \approx 
    \langle{\rm IPR}_{T}^{A}\rangle\langle\sigma_{A}^{2}\rangle+(1-\langle{\rm IPR}_{T}^{A}\rangle)(\sigma_{\rm ex}^{2}-\frac{1}{N}\langle\sigma^{2}_A\rangle)\frac{N}{N-1}
    \label{eq:sigm_t_2nd_step}
\end{equation}
Then, for large $N \gg 1$, keeping only terms of order $\sigma^2_{\rm ex} \sim N^2$, we obtain
\begin{equation}
   \langle\sigma_{T}^{2}\rangle \approx 
    \langle{\rm IPR}_{T}^{A}\rangle\langle\sigma_{A}^{2}\rangle+(1-\langle{\rm IPR}_{T}^{A}\rangle) \sigma_{\rm ex}^{2}
 \label{eq:sigm_t_3rd_step}
\end{equation}

\twocolumngrid
\bibliography{references}

\end{document}